# SEARCH of PRIMARY COSMIC RAYS SOURCES at $5\cdot 10^{13} - 5\cdot 10^{14}$ eV with TIEN SHAN CHRONOTRON - KLARA ARRAY

E.N. Gudkova, N.M. Nesterova.

*P.N. Lebedev Physical institute of Russian Academy (FIAN). Moscow.119991, Moscow,*

*Leninsky pr. 53, FIAN, OYAFA, OKI.*

## ABSTRACT

The primary cosmic ray sources are searched by means of CHRONOTRON - KLARA separate array of the P.N. Lebedev Physical Institute Tien Shan station. It was done on the base of 35 millions observed PCR extensive air showers from $5\cdot 10^{13}$ to $5\cdot 10^{14}$ eV energies. The data analysis was carried on the method of the direct selection of local areas in equatorial coordinates where the deviation of event numbers exceeded the definite value from normal Gaussian standard. These directions are compared with other arrays observed results and with coordinates of astrophysical sources.

Keywords: cosmic rays, extensive air shower, anisotropy

## INTRODUCTION

The aim of this work was to search local sources of primary cosmic rays (PCR) at $E_0 = 5\cdot 10^{13} - 5\cdot 10^{14}$ eV energies according to the HRONOTRON-KLARA array data. The experiments were performed by P.N. Lebedev Physical Institute (FIAN) of Russian Academy of Sciences in collaboration with the Central Research Institute for Physics (KFKI) of Hungarian Academy of Sciences on the separate array at Tien Shan high altitude scientific station (43.04° N, 76.93° E, P = 690 g cm$^{-2}$). The array was intended for continuous investigations of the PCR anisotropy [1, 2, 3]. It operated simultaneously but independently from the FIAN large SHAL installation [4, 5]. The initial experimental data bank contained 35 million extensive air showers (EAS).

The main other experimental investigations of EAS anisotropy currently deal with energies lower than 10 TeV or higher than $10^{19}$ eV. The energy interval that we investigated was studied poorly due to the low intensity of EAS events. Most arrays recording the above-ground characteristics of EAS operated at the lower boundary of our measurements. MILAGRO, IceCube, KASCADE, Tibet-ASγ and other experiments sought elevated fluxes of primary cosmic rays. The contribution from PCR gamma-quanta was evaluated with Tien Shan main complex SHAL installation at $E_0 = 5\cdot 10^{14}$ eV. EAS without muons and hadrons were selected to identify the share of PCR gamma sources for this purpose [6, 7]. The SHAL installation contained electron, muon and hadron detectors. Hadrons were recorded by means of the ionizing calorimeter [4]. It was found that the contribution from primary gamma was $3.3\cdot 10^{-3}$ from all EAS at $E_0 \approx 5\cdot 10^{14}$ eV. The KASCADE experiment [8] used the similar criterion to identify the share of gamma - PCR (EAS selections without muons and "hadrons") in the same energy range (PeV). Events with no hadrons were selected according to parameters S of lateral distributions of electron but without using the KASCADE calorimeter. This way was less definite and reliable. KASCADE experiment revealed no directions with notably elevated EAS fluxes. The upper boundary of anisotropy was estimated as $10^{-3}$ at 0.7 PeV PCR [8]. It is interesting to compare our results with observations at somewhat lower energies.

## RECORDING AND PROCESSING

FIAN Tien Shan CHRONOTRON-KLARA array was consisted of two parts: CHRONOTRON and KLARA. The CHRONOTRON determined the direction of the EAS axis inclination in the horizontal coordinate system (zenith angle θ and azimuth angle φ). The KLARA system recorded the CHRONOTRON reliable readings, the EAS arrival time t and the number of particles in the electron detectors of every EAS [1]. This work present data with all $E_0$ values. Events with θ = 20°-60° are selected for further processing based on details of data registration. The accuracy of θ and φ angles measuring was no more than 3°. After transition from θ, φ, t to the equatorial coordinate system: (the right ascension α, the declination δ, the sidereal time $t_S$) we obtained distribution map according to the EAS number in cells α × δ = 7.5° × 5°. 23 millions of EAS (62% of the initial number) were subjected to final processing (mainly due to selecting events with θ = 20°-60°). The data of every cell were normalized to the visibility time and to the experimental received dependency of EAS absorption in the atmosphere from zenith angle θ. To confirm the results, the same map was obtained for cells α × δ = 10° × 10°.

For each interval we calculated the standard deviation from the mean number of EAS in a cell: α × δ = 7.5° × 5°. We considered the distribution errors resulting after normalization. In Fig.1 the blue dotted line shows the distribution of the deviation of the EAS number in the cell from the mean value before normalization, approximated by the Gaussian function. The pink solid line shows approach of the same distribution received after normalization on angles θ. Dispersion of the second distribution is 1.3 times more, than the first. This factor is used to determine the final criterion of σ significance. Figure 2 shows the map of identified distinguished directions, where σ is equal in three ranges: σ > 3.0; 3.0–2.0; and 2.0–1.6. We estimated the distribution variation after application of the normalization over the absorption of particles in the atmosphere as the function of zenith angle θ. Some small number of these cells with deviation from "normal" law exceeded expected number. It is possible they are directions to cosmic rays source.

CONCLUSIONS

We obtained a map of directions with exceeded cosmic ray fluxes by analyzing the data of PCR energies from $5\cdot10^{13}$ to $5\cdot10^{14}$ eV at the FIAN Tien Shan station CHRONOTRON - KLARA array. Some of these directions could be directions to PCR sources; some could be caused by fluctuations of the background or inaccuracies of registrations. First of all it is worth paying attention to cells with great σ, for example: α = 0-10°, δ = 5-10°. It should be noted that two pulsars are in this direction.

More reliable directions to sources can be found by means of confirming with results of other arrays experiments. Several directions presented in this work are in an agreement with data obtained at Tien Shan main complex station installation SHAL (the selection of showers with no muons and hadrons from gamma - PCR) (for example: α=90-120°, δ=45-55° and α=215-230°, δ=15-20°). (Fig. 2, red crosses) [6.7]. It is necessary to allocate the excess of the extensive region α=150-230°, δ=20-35°. Here the excess is observed also with SHAL MSU and PRO1000 data [10].

Earlier we compared data of our previous FIAN results [2, 3] with previous data of two arrays of Moscow State University: SHAL MSU for EAS at $E_0 \approx 6\cdot10^{14}$ eV and PRO-1000 [9] for EAS at $E_0 \approx 2\cdot10^{14}$ eV. The agreement of several directions with exceeded cosmic ray fluxes was found.

The same data with FIAN array had been analyzed by the new another method using "shuffling" and formula by Li & Ma [11] in the up-to-date work [10]. The multiple overlapping circles had been using for treatment of data [10], therefore area with elevated EAS fluxes are so massive and statistical significances of the distribution differ from normal Gauss law. Regions of cosmic rays with elevated EAS fluxes were received by two different methods (this work and [10]) basically coincide if to take into account the difference of data statistical significance coefficients. It is possible directions of excess fluxes reflect a real PCR anisotropy, but also they can be random fluctuations of the background or errors of measurements. Still, it is interesting that some of the EAS excess regions are located at the same regions of the celestial sphere in three data sets: FIAN, SHAL MSU and PRO1000.

We also see a good agreement between our data and results obtained with the MILAGRO [13] and IceCube [14] arrays. Areas of elevated PCR fluxes identified with the MILAGRO experiment (rectangular areas A and B) are represented in Figure 2. Several chosen cells are located in the Crab and Geminga region. The extensive (~20° in the diameter) Galactic object there is in this region. It is the Monogem Ring, the supernova's residual star. The young pulsar PSR B0656+14 was observed in soft gamma ray is near the center of this area [15]. Can this pulsar is the source of PCR at energies above $5 \cdot 10^{13}$ eV?

Probably it is not possible to observe an excess of charged particles from energies pulsars and residuals of supernova stars since these particles lose their initial direction by twisting in the Galactic magnetic fields. But now there are new models usable to develop ways to record them. One of up-to-date theory is that there are Galactic magnetic fields with unique configurations (magnetic lenses). Unfortunately same energies neutrons are unable to reach Earth because of the short free paths. Possible sources of primary cosmic rays at $5 \cdot 10^{13} - 5 \cdot 10^{14}$ energies are gamma - sources.

## ACKNOWLEDGMENTS

The authors would like to thank the colleagues of KFKI and the Tien Shan station for engineering of array, making our measurements and proceedings.

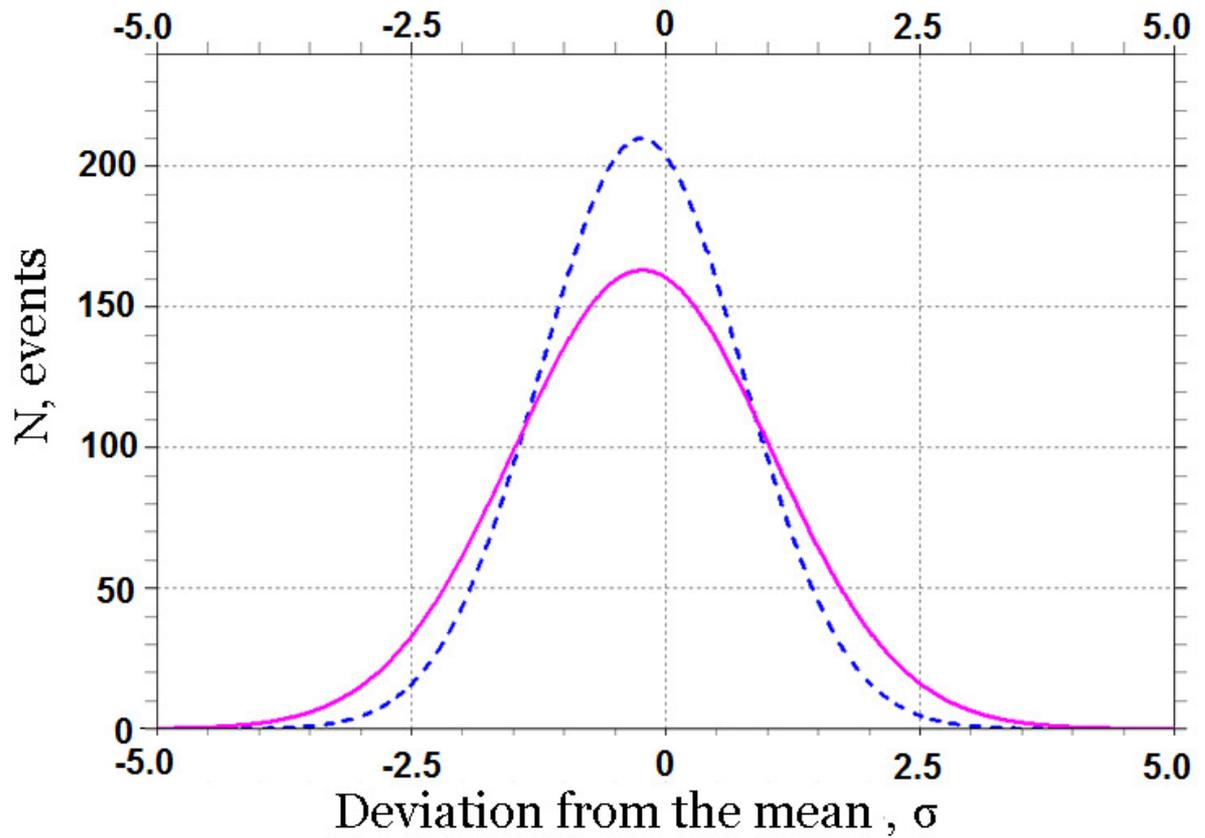

**Fig.1.** Distribution of the deviation of shower numbers in a cell α, δ from mean σ value. The experimental curve for the initial data is approximated by a Gaussian function and it is shown by the blue dashed line; $D = 0.933$. The pink solid line is our approximation of the same distribution obtained after normalization over the zenith angle; $D = 1.27$.

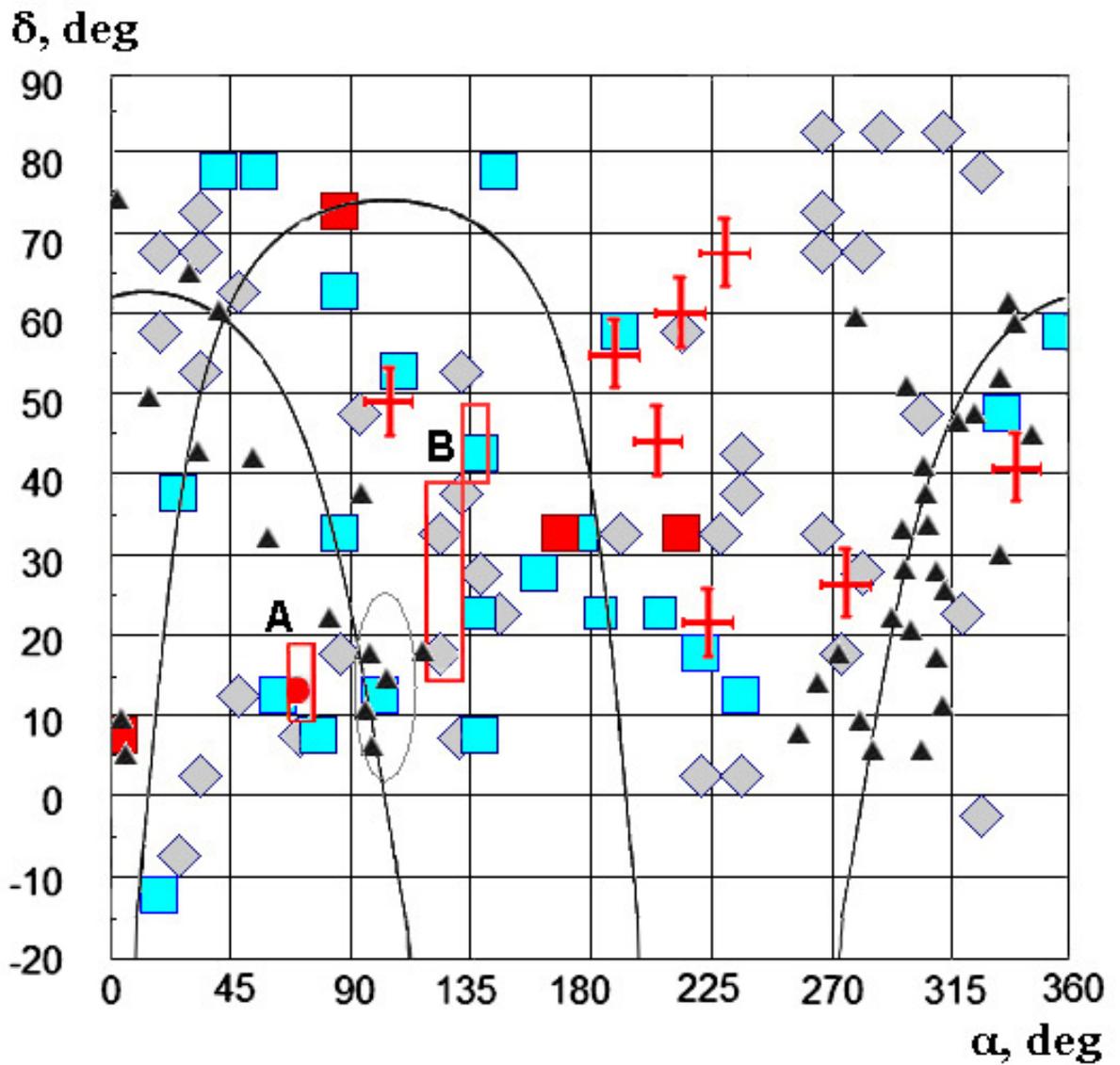

**Fig.2.** Directions in the equatorial coordinates with standard deviations σ: > 3σ - red rectangles; (σ = 2.1 – 3.0) - blue rectangles; and (σ =1.6 – 2.0) – gray diamonds. The red crosses denote are coordinates of showers free of muons and hadrons from gamma-PCR [6, 7]. The two rectangles and points A and B are the MILAGRO data [13]. Black triangles denote gamma-pulsars, the oval - Monogem Ring. Curves show the Galactic and the Supergalactic planes.